\documentclass[ aip,
reprint, twocolumn 
]{revtex4-1}

\usepackage{graphicx}
\usepackage{dcolumn}
\usepackage{bm}
\usepackage[english]{babel}
\usepackage{amssymb,amsmath}
\usepackage{mathtools}
\usepackage{physics}
\usepackage{float}
\usepackage{longtable} 
\usepackage[caption=false]{subfig}
\usepackage{hyperref}
\usepackage[utf8]{inputenc} 
\usepackage[T1]{fontenc}   
\usepackage{textcomp}      
\usepackage{multirow}
\usepackage[utf8]{inputenc}
\usepackage[T1]{fontenc}
\usepackage{mathptmx}
\usepackage{etoolbox}

\makeatletter
\def\@email#1#2{%
 \endgroup
 \patchcmd{\titleblock@produce}
  {\frontmatter@RRAPformat}
  {\frontmatter@RRAPformat{\produce@RRAP{*#1\href{mailto:#2}{#2}}}\frontmatter@RRAPformat}
  {}{}
}%
\makeatother
\begin{document}

\preprint{AIP/123-QED}

\title{Predicting the Curie temperature in substitutionally disordered alloys using a first-principles based model}
\author{Marian Arale Brännvall}
\author{Rickard Armiento}%

\author{Björn Alling}
\affiliation{Department of Physics, Chemistry and Biology, Linköping University, 581 83, Linköping, Sweden}
\date{\today}

\begin{abstract}
When exploring new magnetic materials, the effect of alloying plays a crucial role for numerous properties. By altering the alloy composition, it is possible to tailor, e.g., the Curie temperature ($T_\text{C}$). In this work, $T_\text{C}$ of various alloys is investigated using a previously developed technique [\href{https://doi.org/10.1103/PhysRevMaterials.8.114417}{Br\"{a}nnvall \textit{et al.}\ Phys. Rev. Mat. (2024)}] designed for robust predictions of $T_\text{C}$ across diverse chemistries and structures. The technique is based on density functional theory calculations and utilizes the energy difference between the magnetic ground state and the magnetically disordered paramagnetic state. It also accounts for the magnetic entropy in the paramagnetic state and the number of nearest magnetic neighbors. The experimentally known systems, Fe$_{1-x}$Co$_x$, Fe$_{1-x}$Cr$_x$, Fe$_{1-x}$V$_x$, NiMnSb-based Heusler alloys, Ti$_{1-x}$Cr$_x$N, and Co$_{1-x}$Al$_x$ are investigated. The experimentally unexplored system Fe$_{1-x}$Tc$_x$ is also tested to demonstrate the usefulness of the developed method in guiding future experimental efforts. This work demonstrates the broad applicability of the developed method across various systems, requiring less hands-on adjustments compared to other theoretical approaches. 
\end{abstract}

\maketitle

\section{Introduction}
\label{sec:intro}

Materials with a spontaneous macroscopic magnetization lose it at the Curie temperature ($T_\text{C}$) due to a transition from a ferro- or ferrimagnetically ordered state to a disordered paramagnetic state. In the search for technologically applicable magnets, knowledge of this temperature is crucial. For magnetic alloys, the effect of different compositions is of high interest as it enables tailoring $T_\text{C}$ of a magnetic material. Conducting reliable predictions of $T_\text{C}$ in general and with respect to the composition of an alloy in particular, is therefore of great importance. In alloy systems, changes in composition can modify both the number and strength of magnetic exchange interactions, and the magnetic ordering itself. For example, increasing the concentration of a nonmagnetic element reduces the number of magnetic neighbors, weakening the overall magnetic coupling and lowering $T_\text{C}$. Conversely, alloying with elements that possess strong magnetic moments—such as transition metals with partially filled $d$ or $f$ orbitals—can enhance the magnetic interactions within an alloy. These elements contribute to stronger exchange interactions, which stabilize magnetic order at higher temperatures, resulting in an increased Curie temperature~\cite{haglund_curie_1982}.

The Curie temperature can be measured by means of experiments and derived using theoretical calculations.  The experimental approach quickly becomes too time-consuming and costly to apply when exploring a wide range of new magnetic materials. In theoretical calculations, one may, for example, construct a Heisenberg Hamiltonian and derive the exchange interactions between magnetic moments using the method developed by Liechtenstein \textit{et al.}~\cite{liechtenstein_local_1987}, subsequently determining $T_\text{C}$ using Monte Carlo simulations or a mean-field approximation \cite{rusz_exchange_2006,lezaic_first-principles_2007,alling_effect_2009, alling_theory_2010}. This method is, however, often impractical for large-scale investigations of systems with different structures and chemistries, as it requires extensive human manual input. Hence, studies employing this method often focus on only one or a few similar systems \cite{halilov_adiabatic_1998, kubler_ab_2006, rusz_exchange_2006,rusz_ab_2006,alling_effect_2009,alling_theory_2010, zelai_first_2023}. 

To accelerate theoretical predictions of the Curie temperature, machine learning (ML) methods have been introduced in multiple studies \cite{nelson_predicting_2019, belot_machine_2023, dam_important_2018, long_accelerating_2021, lu_--fly_2022,nguyen_ensemble_2019, xu_predicting_2024, hu_searching_2020,zhai_accelerated_2018}. This approach is highly compatible with a high-throughput search of magnetic materials. However, attempts at developing general ML models for predicting the Curie temperature have faced challenges in capturing a consistently physical and continuous variation of $T_\text{C}$ as a function of composition \cite{nelson_predicting_2019, belot_machine_2023}.

As mentioned previously, when the composition of an alloy changes, so does the number and type of nearest magnetic neighbors, leading to variations in exchange interactions and the stability of the magnetic ground state. Understanding and incorporating the effects of these factors is crucial for predicting how alloying will influence the magnetic properties of materials. To tackle these systems it therefore appears particularly important to employ physics-based predictive models. In a previous work~\cite{brannvall_predicting_2024}, we created a physics-based model, utilizing output from density functional theory (DFT) calculations, and found it to greatly improve the predictions of the substitutionally disordered alloy Fe$_{1-x}$Co$_x$ compared to ML models. The model uses the energy difference between the paramagnetic state and the magnetic ground state, the magnetic entropy, and the number of nearest magnetic neighbors. The paramagnetic state is simulated using a supercell implementation of the disordered local moments (DLM) method \cite{alling_effect_2010, gyorffy_first-principles_1985}, and the magnetic ground state is determined by a ground-state search method introduced in the work by Ehn \textit{et al.}\ \cite{ehn_first-principles_2023}. Details of the DLM simulations and ground-state search are described in Sec. \ref{sec:method_DFT_calculations}. In the current work, we explore the applicability of this model further by applying it to estimate $T_\text{C}$ of the  disordered alloys Fe$_{1-x}$Co$_x$, Fe$_{1-x}$Cr$_x$, Fe$_{1-x}$V$_x$, Ni$_x$Cu$_{1-x}$MnSb, Ni$_x$MnSb, Ti$_{1-x}$Cr$_x$N, and Co$_{1-x}$Al$_x$. The results are compared to the experimental $T_\text{C}$. Additionally, the model is applied to Fe$_{1-x}$Tc$_x$ to assess its predictive accuracy in a system where experimental data is scarce due to the radioactive nature of technetium. This allows for theoretical insights that may be useful to guide future experimental validation efforts.  We also address and analyze some limitations of our model, discussing their underlying causes. 

This current investigation shows that the method from Ref.~\cite{brannvall_predicting_2024} can be applied across a wide range of systems, unlike other theoretical methods that often require  hands-on adjustments for each specific case. Overall, it offers a reliable approach for predicting and tailoring the Curie temperature by adjusting the alloy composition.


\section{Method}
\subsection{Model Description}
In Ref.~\cite{brannvall_predicting_2024}, four models were developed to predict the Curie temperature of magnetic materials based on output from DFT calculations. The most elaborate model has the form,
\begin{equation}\label{eq:model_D_form}
    T_\text{C}= A\frac{\Delta E }{S^\text{mag}} \Bigg ( 1 - \frac{B}{NN^{C}}\Bigg ) + D \: \text{K},
\end{equation}
where the parameters $A$, $B$, $C$, and $D$ are fitted to the 32 known ferro- and ferrimagnetic materials presented in Ref.~\cite{brannvall_predicting_2024} giving $A=0.85$. $B=0.69$, $C=0.14$, and $D=124$; 
 $\Delta E = E_\text{DLM} - E_\text{GS}$ is the energy difference between the DLM state and the magnetic ground state found after the ground-state search (described more in Section \ref{sec:method_DFT_calculations}); $NN$ is the number of nearest magnetic neighbors of the magnetic atoms, defined as atoms with magnetic moments larger than 0.75 $\mu_B$ in the magnetic ground state; $S^\text{mag}$ is the magnetic entropy of the DLM state given as,
 \begin{equation}
\label{eq:S_mag}
    S^\text{mag} = k_\text{B}\Bigg[\sum_{i=1}^{N_\text{mag}}\ln{(m_i + 1)}\Bigg]/N_\text{mag},
\end{equation}
where $k_\text{B}$ is the Boltzmann constant, $N_\text{mag}$ is the number of magnetic atoms with constrained magnetic moments, and $m_i$ is the magnetic moment magnitude of the $i$th constrained magnetic moment.

None of the alloys investigated in the present work are included among the 32 systems used to obtain the fitting parameters of Eq.~(\ref{eq:model_D_form}) and can thus be seen as a critical test of its generality. The parameters $B$ and $C$ are optimized using the Nelder-Mead method \cite{nelder_simplex_1965}. A linear fit to the experimental $T_\text{C}$ subsequently gives the parameters $A$ and $D$ \cite{brannvall_predicting_2024}.

\subsection{Rationale Behind the Model Formulation \label{sec:rationale_model}}
The motivation for the form of Eq.~(\ref{eq:model_D_form}) is given in more detail in Ref.~\cite{brannvall_predicting_2024}, and is briefly outlined here. There are three physical quantities in Eq.~(\ref{eq:model_D_form}): $\Delta E$, $S^\text{mag}$, and $NN$. Through a mean-field treatment of the magnetic system an estimate of $T_\text{C}$ can be expressed in terms of the energy difference between the paramagnetic state and the ferromagnetic ground state \cite{turek_ab_2003, liechtenstein_magnetic_1983, bergqvist_theoretical_2007}.
The connection between the magnetic entropy, $S^\text{mag}$, and $T_\text{C}$ can be argued via the approximation of the free energy per magnetic atom,
\begin{equation}
    F_\text{DLM} = E_\text{DLM} - k_BT\ln{(m + 1)} = E_\text{DLM} - S^\text{mag}T,
\end{equation}
where $E_\text{DLM}$ is the energy of the DLM state and $m = \mu /\mu_B$ is the atomic magnetic moment in the
magnetically disordered state; $\mu$ is the  magnitude of the atomic magnetic
moment given in Bohr magneton units, $\mu_B$. The Curie temperature can then be approximated as
\begin{equation}
\label{eq:T_C_Smag}
    T_\text{C} \propto (E_\text{DLM} - E_\text{FM}) / S^\text{mag},
\end{equation}
assuming that the ordered magnetic state remains ideally ordered and the paramagnetic state is ideally disordered, thus neglecting all effects of short-range order (SRO). These SRO effects can be shown to vary in importance depending on whether the magnetic energy is distributed over many or few magnetic nearest neighbors. Based on the findings presented in Ref.~\cite{brannvall_predicting_2024}, the expression for $T_\text{C}$ in Eq.~(\ref{eq:T_C_Smag}) can be adjusted with a factor $1 - \frac{B}{NN^C}$
to capture the different impact of SRO effects in different crystal structures and alloy compositions. This adjustment results in a lower estimated $T_\text{C}$ for a given energy difference when the number of nearest neighbors is small. The parameters B and C are the same as in Eq.~(\ref{eq:model_D_form}).

In disordered alloys, the number of magnetic neighbors varies with composition. Introducing a certain percentage of a nonmagnetic component reduces the number of magnetic neighbors per magnetic atom by approximately the same fraction. Our model accounts for this effect on the value of $NN$, adjusting the Curie temperature in response not only to different coordination numbers of different crystal structures but also to varying proportions of magnetic and nonmagnetic elements in the alloy.

\subsection{Density Functional Theory Calculations \label{sec:method_DFT_calculations}}

To obtain the energies of the paramagnetic and ground states, the disordered local moments method and a ground-state search are used, respectively. The ground-state search is performed using the method presented by Ehn \textit{et al.}\ \cite{ehn_first-principles_2023} and described in Ref.~\cite{brannvall_predicting_2024}. This approach involves running noncollinear magnetic calculations with initial random directions for the magnetic moments. During ionic relaxation, the directions are free to rotate towards their energetically preferred orientations and adjust in magnitude. To simulate substitutionally disordered alloys, the special quasirandom structure  method (SQS) is used \cite{zunger_special_1990}. We perform a single ground-state search with these SQS cells and compare the total energy to that of a collinear ferromagnetic configuration. In the investigated cases, the energies were found to be comparable, suggesting that a ferromagnetic ground state is the correct ground state, consistent with experimental expectations for these systems.

The DLM method used here is based on the supercell approach implemented by Alling \textit{et al.}\ \cite{alling_effect_2010}. In the present work, the magnetic moments are set with random directions (in a noncollinear configuration) and with magnetic moment sizes based on the average sizes of each element from the ground-state search. Only one DLM configuration is used to limit the number of computationally intensive calculations. To ensure a nearly ideal DLM configuration, the directions are set so that the SRO is as close to zero as possible. The SRO is defined as,
\begin{equation}
    \mathrm{SRO} = \Bigg \langle \frac{1}{N_{\text{at}}} \sum_{i=1}^{N_{\text{at}}} \frac{1}{N_{\text{NN}}}\sum_{j=1}^{N_{NN}}\pmb{e}_i\cdot\pmb{e}_j \Bigg\rangle,
\end{equation}
\noindent
 where, $N_{\text{at}}$ is the total number of atoms in the cell and $N_{\text{NN}}$ is the number of magnetic neighbors in the first coordination shell. In the DLM calculation, the directions are constrained using the method by Ma and Dudarev \cite{ma_constrained_2015}, where the direction and sign of the magnetic moment are fixed, preventing rotation or flips, while allowing its magnitude to vary. Only elements with magnetic moment magnitudes above 0.75 $\mu_B$ in the ground-state search are constrained in the DLM run. This limit must, however, be altered in some of the investigated cases; this will be discussed in Sec.~\ref{sec:discussion}. The constraining parameter $\lambda$, which determines the strength of the penalty energy, is set incrementally until the slope, given by
\begin{equation}
    \frac{\abs{E_{new} - E_{old}}}{\abs{\lambda_{new} - \lambda_{old}}}, 
\end{equation}
is less than 0.1 meV.

From the DLM calculation, the magnetic moment magnitudes required for calculating the magnetic entropy, $S^\text{mag}$, in Eq.~(\ref{eq:S_mag}) are obtained as the sum of the spin projections into the orbitals within the projector augmented-wave potential sphere. 
\section{Computational Details}
To obtain the energies and the magnetic moment sizes, we use DFT calculations as implemented in the Vienna Ab initio Simulation Package (VASP) \cite{kresse_ab_1993, kresse_ab_1994, kresse_efficiency_1996, kresse_efficient_1996} using  projector augmented-wave potentials (PAW) \cite{blochl_projector_1994,kresse_ultrasoft_1999} and the Perdew-Burke-Ernzerhof (PBE) generalized gradient approximation (GGA) for approximating the exchange-correlation functional \cite{perdew_generalized_1996}. The energy cutoff is set to 450 eV. The chemical disorder in the investigated alloys is modeled using the special quasi-random structure (SQS) method \cite{zunger_special_1990}. For iron-containing alloys, 64-atom SQS supercells are used, except for compositions with 5\% or 10\% of non-iron elements, where 125-atom supercells are employed. Supercells for $\mathrm{Ti_{1-x}Cr_x N}$ contain 128 atoms, while $\mathrm{Ni_x Cu_{1-x}MnSb}$ is modeled using 96-atom supercells. In the case of $\mathrm{Ni_x MnSb}$, the supercell size increases with the nickel content. Starting from a 96-atom NiMnSb supercell, additional nickel atoms are added as needed. For example, at $x=1.5$, 16 extra nickel atoms are included, resulting in a total of 112 atoms in the supercell. A Monkhorst-Pack 2$\times$2$\times$2 k-point grid is used for all substitutionally disordered alloys \cite{monkhorst_special_1976}.

To simulate the paramagnetic state, the DLM approach is used \cite{alling_effect_2010, gyorffy_first-principles_1985}. The calculations are spin-polarized, utilizing noncollinear magnetism. The magnetic moments are initially set to 3 Bohr magneton ($\mu_B$) in random directions in the ground-state search. The magnetic moments are then allowed to relax in both magnitude and direction. In the DLM simulations, the directions of the magnetic moments are constrained using the method presented in Ref.~\cite{ma_constrained_2015}.  The magnetic moment magnitudes are allowed to relax in the DLM run, but are initially set to 110\% of the relaxed average ground-state size of the magnetic moments of the specific atomic species.  
\section{Results}
\subsection{Prediction of \textit{T}$_\text{C}$ Across Compositions}

\begin{figure*}
    \centering
    \includegraphics[scale=0.45]{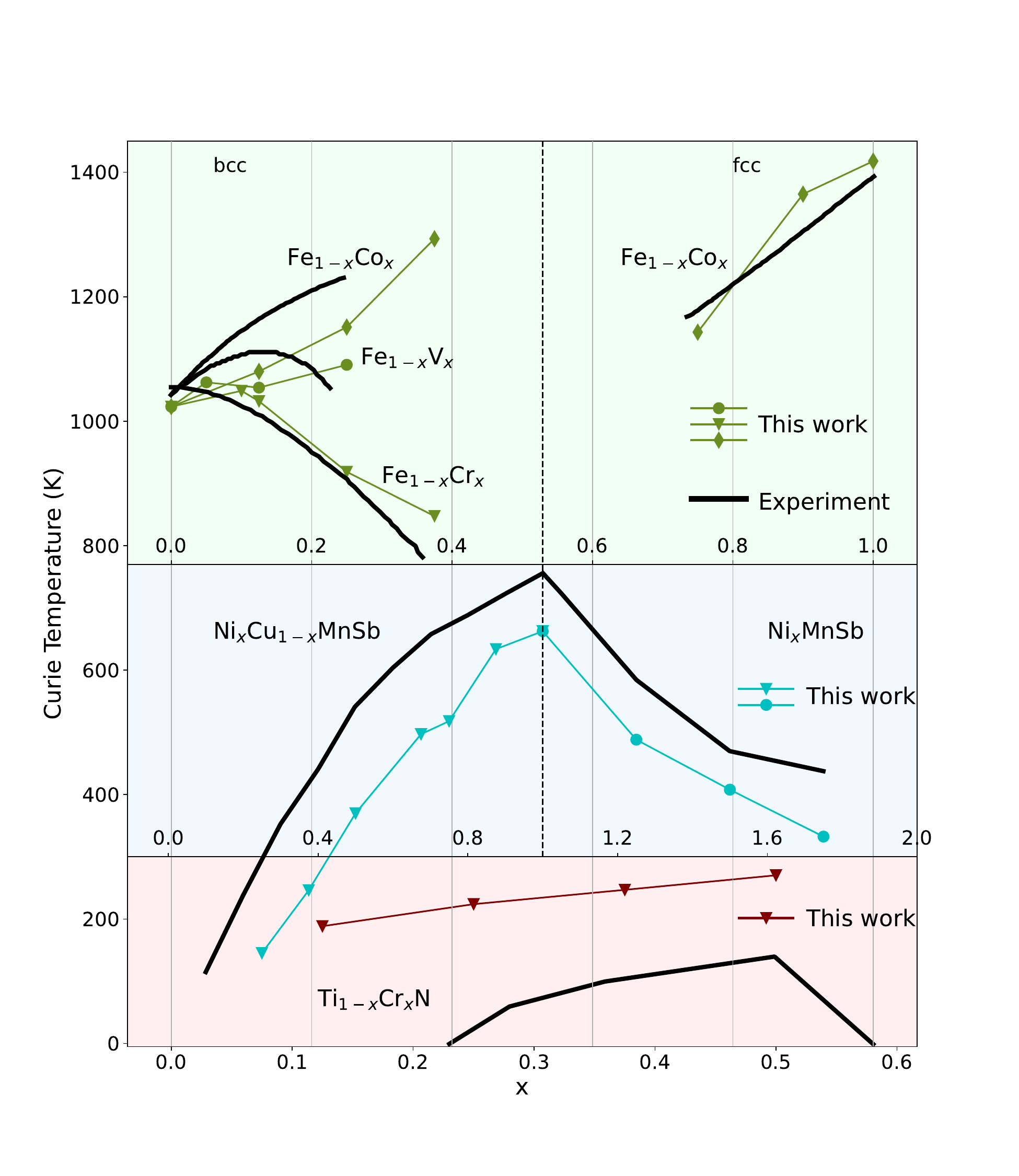}
    \caption{Predicted and experimental Curie temperatures as a function of composition are shown for the following alloy systems: bcc and fcc Fe$_{1-x}$Co$_x$ (experimental data from \cite{nishizawa_cofe_1984, 69Stu, 75Nor}), bcc Fe$_{1-x}$V$_x$ (experimental data from \cite{smith_fev_1984, 09Por, 10Hon, 25Mau, 30Oya, 30Wev, 31Vog, 34Abr, 36Fal}), bcc Fe$_{1-x}$Cr$_x$ (experimental data from \cite{wijn_3d_1997}), Ni$_x$Cu$_{1-x}$MnSb (experimental data from \cite{ren_magnetic_2005}), Ni$_x$MnSb (experimental data from \cite{neibecker_atomic_2022}), and Ti$_{1-x}$Cr$_x$N (experimental data from \cite{inumaru_ferromagnetic_2007}). Here, $x$ represents the fractional substitution of one element for another in each alloy, except for Ni$_x$MnSb, where $x$ denotes the Ni concentration, varying from 1 to 2. The numerical values of this figure are found in Table \ref{tab:alloys_numerical_values}.}
    \label{fig:all_alloys_composition}
\end{figure*}

Seven different substitutional random alloy systems are investigated using the model of Eq.~(\ref{eq:model_D_form}) and compared with experimental results. Fig.~\ref{fig:all_alloys_composition} shows the results of predicted $T_\text{C}$ as a function of composition for  body-centered cubic (bcc) and face-centered cubic (fcc) Fe$_{1-x}$Co$_x$, bcc Fe$_{1-x}$V$_x$, bcc Fe$_{1-x}$Cr$_x$, Ni$_x$Cu$_{1-x}$MnSb, Ni$_x$MnSb, and Ti$_{1-x}$Cr$_x$N. The numerical values are found in Table~\ref{tab:alloys_numerical_values}. The different Fe-containing alloys in the upper part of Fig.~\ref{fig:all_alloys_composition} are mainly of the bcc structure. However, for Fe$_{1-x}$Co$_x$ there is a phase transition from bcc to fcc coinciding with the magnetic transition between 31 and 78 at \% iron in the cobalt- and iron-rich regions, respectively \cite{nishizawa_cofe_1984}. Hence, for the cobalt-rich region the random alloy is of fcc structure. The experimental results for the Fe$_{1-x}$Co$_x$ system are from Ref.~\cite{nishizawa_cofe_1984}, which compiled the data from Refs.~\cite{69Stu, 75Nor}. The experimental results for Fe$_{1-x}$V$_x$ are from Ref.~\cite{smith_fev_1984}, which is a collection of results from several experimental studies \cite{09Por, 10Hon, 25Mau, 30Oya, 30Wev, 31Vog, 34Abr,36Fal}. The experimental Curie temperatures for the Fe$_{1-x}$Cr$_x$ are from the Ref.~\cite{wijn_3d_1997}. 

In the Ni$_x$MnSb system, $x$ represents the concentration of Ni, ranging from 1 to 2, indicating a variation in the Ni content from half to full Heusler alloys, i.e., NiMnSb to Ni$_2$MnSb. In contrast, in the Ni$_x$Cu$_{1-x}$MnSb system, $x$ is the fractional substitution of Ni for Cu, ranging from 0 to 1. The structure of the Heusler alloy Ni$_2$MnSb has been found in earlier investigation to be of the L2$_1$ type, whereas the half Heusler alloy NiMnSb has a C1$_b$ structure \cite{webster_magnetic_1968, szytula_atomic_1972}. Webster and Mankikar found that from NiMnSb to Ni$_{1.6}$MnSb the structure is ordered in the C1$_b$ structure. However, at higher nickel concentrations, there is significant disorder from this structure \cite{webster_chemical_1984}. The investigated compositions in Fig.~\ref{fig:all_alloys_composition} for Ni$_x$MnSb are all in the C1$_b$ structure. The experimental $T_\text{C}$ for Ni$_x$Cu$_{1-x}$MnSb are from the work by Ren \textit{et al.}~\cite{ren_magnetic_2005}. For Ni$_x$MnSb, the experimental results are taken from the recent work of Neibecker \textit{et al.}~\cite{neibecker_atomic_2022}.


Ferromagnetism was discovered experimentally in Ti$_{1-x}$Cr$_x$N by Refs.~\cite{aivazov_magnetic_1975, inumaru_ferromagnetic_2007} at the TiN-rich regime. In Ref.~\cite{alling_theory_2010}, this system was investigated using first-principles calculations, confirming the experimental findings. The experimental results in Fig.~\ref{fig:all_alloys_composition} for this system are from Ref.~\cite{inumaru_ferromagnetic_2007}. When conducting the ground-state search, the chromium moments do not end up completely ferromagnetically aligned in any of the cases $x = 0.125, 0.25, 0.375$, or $0.5$. Judging from the total magnetic moment of the supercell, the average magnetic moment is 1.84, 0.79, 0.55, and 1.12 $\mu_B$/chromium atom for the respective $x$-values, reflecting directional variations but no change in magnitude. In the ferromagnetic case, the average chromium moment has a magnitude of approximately $2.2$ $\mu_B$ for all $x$-values. The energies of the ground-state search are compared to a ferromagnetic calculation, and in all cases the ground-state search results in a slightly lower energy. The largest difference is for $x=0.5$, where the ferromagnetic calculation results in an energy 13 meV/chromium atom higher than the ground-state search.


In the workflow developed in our previous work, the magnetic moments to be constrained in the DLM run are determined based on the average magnetic moment magnitude of the specific species obtained in the ground-state search. If the average magnetic moment magnitude  is larger than 0.75 $\mu_B$ in the ferromagnetic ground state, the magnetic moments are constrained; otherwise they are free to change size and direction in the DLM calculations. These atoms are essentially treated on an equal footing with nonmagnetic atoms. Following this strictly in some cases is, however, not possible. For Fe$_{1-x}$Cr$_x$, the chromium moments are on average approximately $1.3$ $\mu_B$ when $x=0.125$ and should therefore be constrained in the DLM run. These moments are, despite this, not constrained to avoid convergence problems, which is discussed in more detail in Sec.~\ref{sec:discussion}. This is also the case for the vanadium moments in Fe$_{1-x}$V$_x$, where, for all tested $x$-values, these moments are larger than $0.75$ $\mu_B$ but are never constrained in the DLM calculations.

\begin{figure}
    \centering
    \includegraphics[scale=0.34]{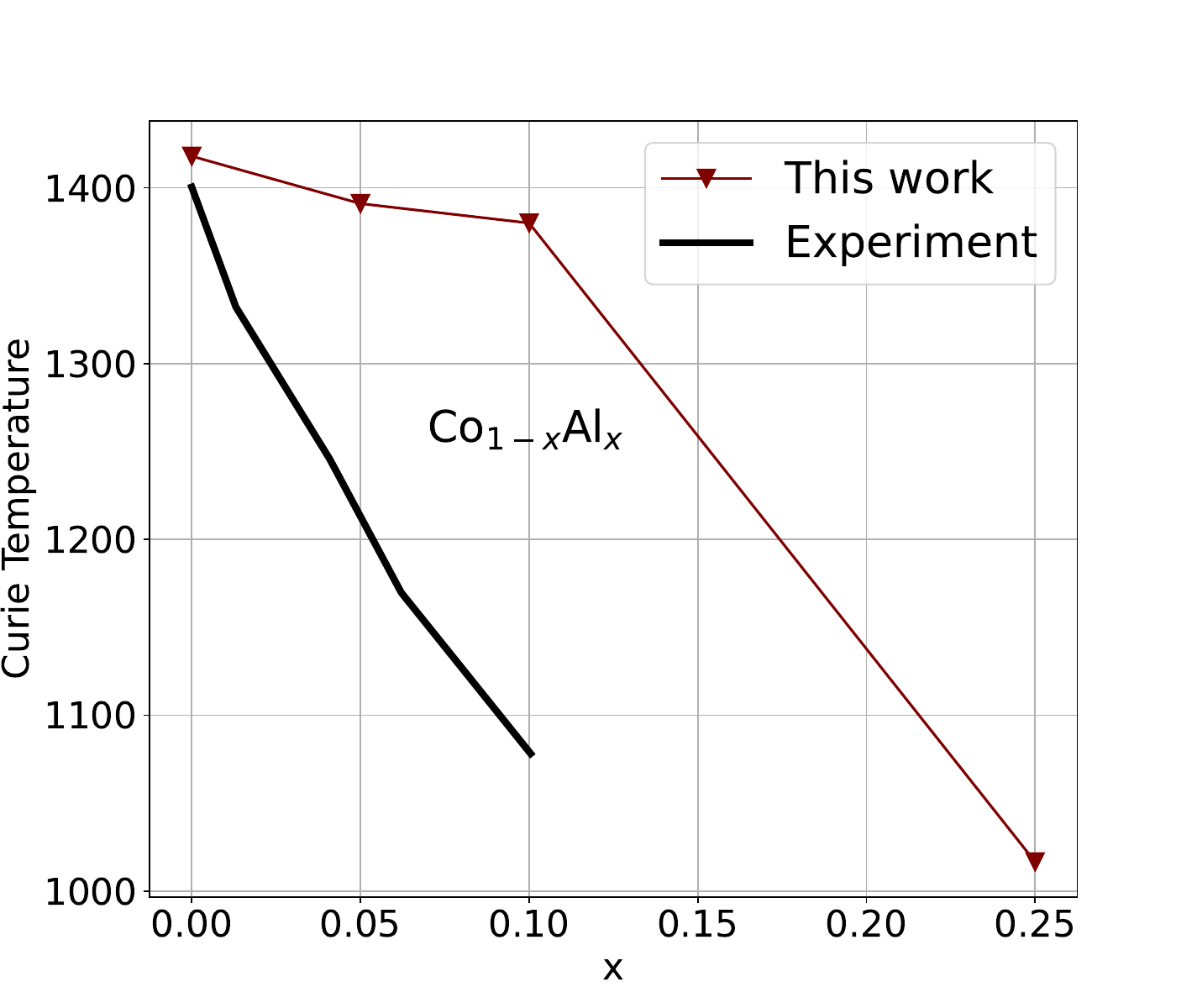}
    \caption{Predicted and experimental Curie temperatures of fcc Co$_{1-x}$Al$_x$ as a function of the fractional substitution of cobalt with aluminum. The experimental Curie temperatures are from Ref.~\cite{mcalister_-co_1989}.}
    \label{fig:CoAl}
\end{figure}

\begin{figure*}
 \subfloat[\label{magmom_Co}]{\includegraphics[scale=0.42]{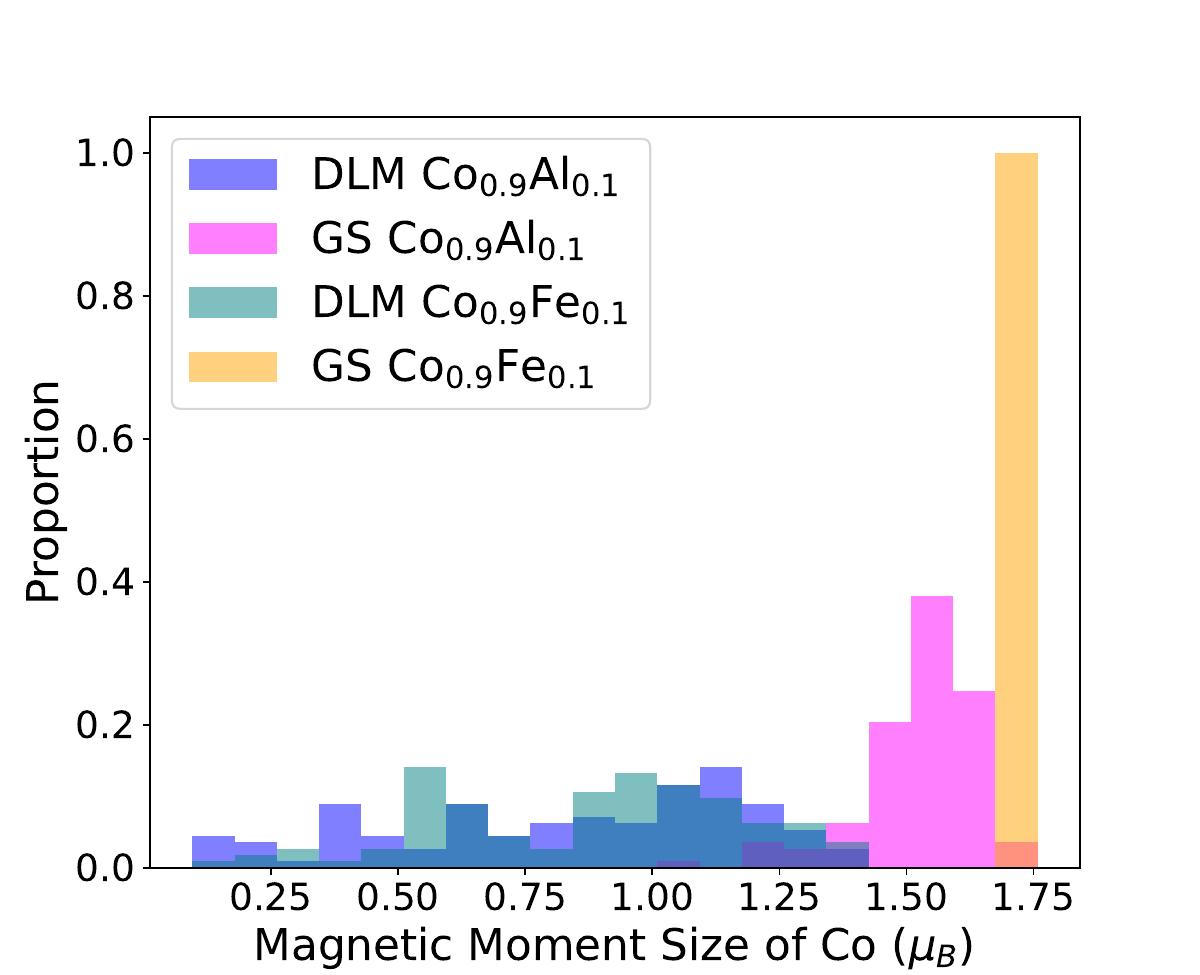}} 
 \subfloat[\label{magmom_Fe-Al}]{\includegraphics[scale=0.42]{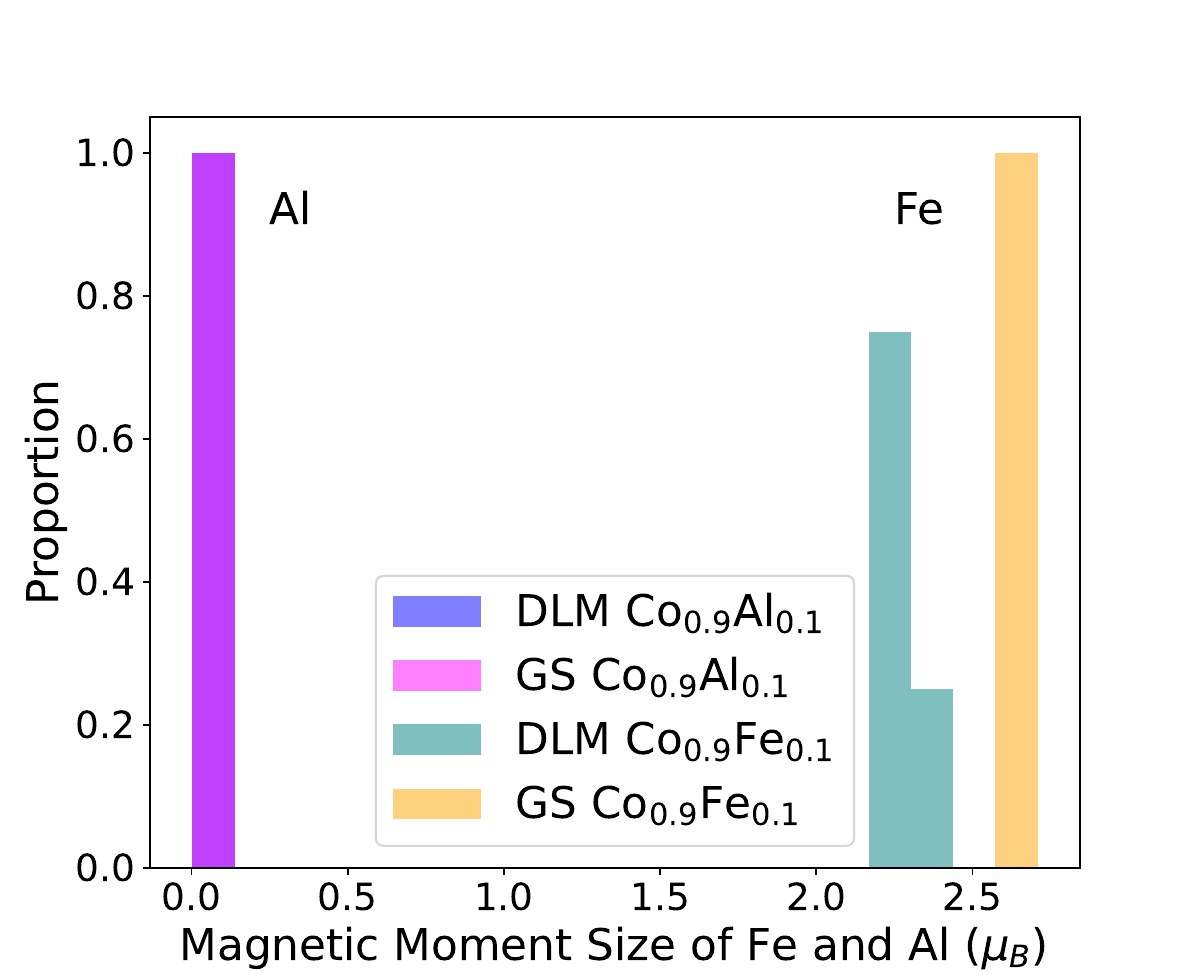}}

\caption{\label{magmom_sizes_CoAl_Co}Distribution of the magnetic moment magnitudes of cobolt, aluminum, and iron in both fcc Co$_{0.9}$Al$_{0.1}$ and Co$_{0.9}$Fe$_{0.1}$. The magnetic moment magnitudes are from the  magnetic ground state (GS) and disordered local moments state (DLM).}
\end{figure*}

The fcc Co$_{1-x}$Al$_x$ system is also investigated using our model. This system is separated from the others as an example of a case with more itinerant type of magnetism to illustrate when our model is not expected to perform as well as in the above examples. This is confirmed in Fig.~\ref{fig:CoAl}. The procedure of increasing the constraining parameter $\lambda$ is slightly adjusted for this alloy system due to a higher sensitivity in the electronic convergence of the self-consistency cycle. After $\lambda=5$ the $\lambda$ parameter is set to 7. In the case of 25\% aluminum, a satisfying level of electronic convergence is not reached for $\lambda=10$. Therefore, Fig.~\ref{fig:CoAl} shows the results of $\lambda=7$ for all percentages of aluminum. The penalty energy in these cases is approximately $0.2$ meV/magnetic atom, which we consider an acceptable level of accuracy.  The experimental values for Co$_{1-x}$Al$_x$ are from Ref.~\cite{mcalister_-co_1989}. In Fig.~\ref{magmom_Co}, the magnetic moment sizes of the cobalt atoms are shown for both Co$_{0.9}$Al$_{0.1}$ and Co$_{0.9}$Fe$_{0.1}$. The moment sizes are given in the DLM state and in the magnetic ground state for comparison. In Fig.~\ref{magmom_Fe-Al}, the magnetic moment sizes of iron and aluminum are given for the same systems. 

\subsection{Prediction of \textit{T}$_\text{C}$ in Fe$_{1-x}$Tc$_x$}
\begin{figure}
    \centering
    \includegraphics[scale=0.34]{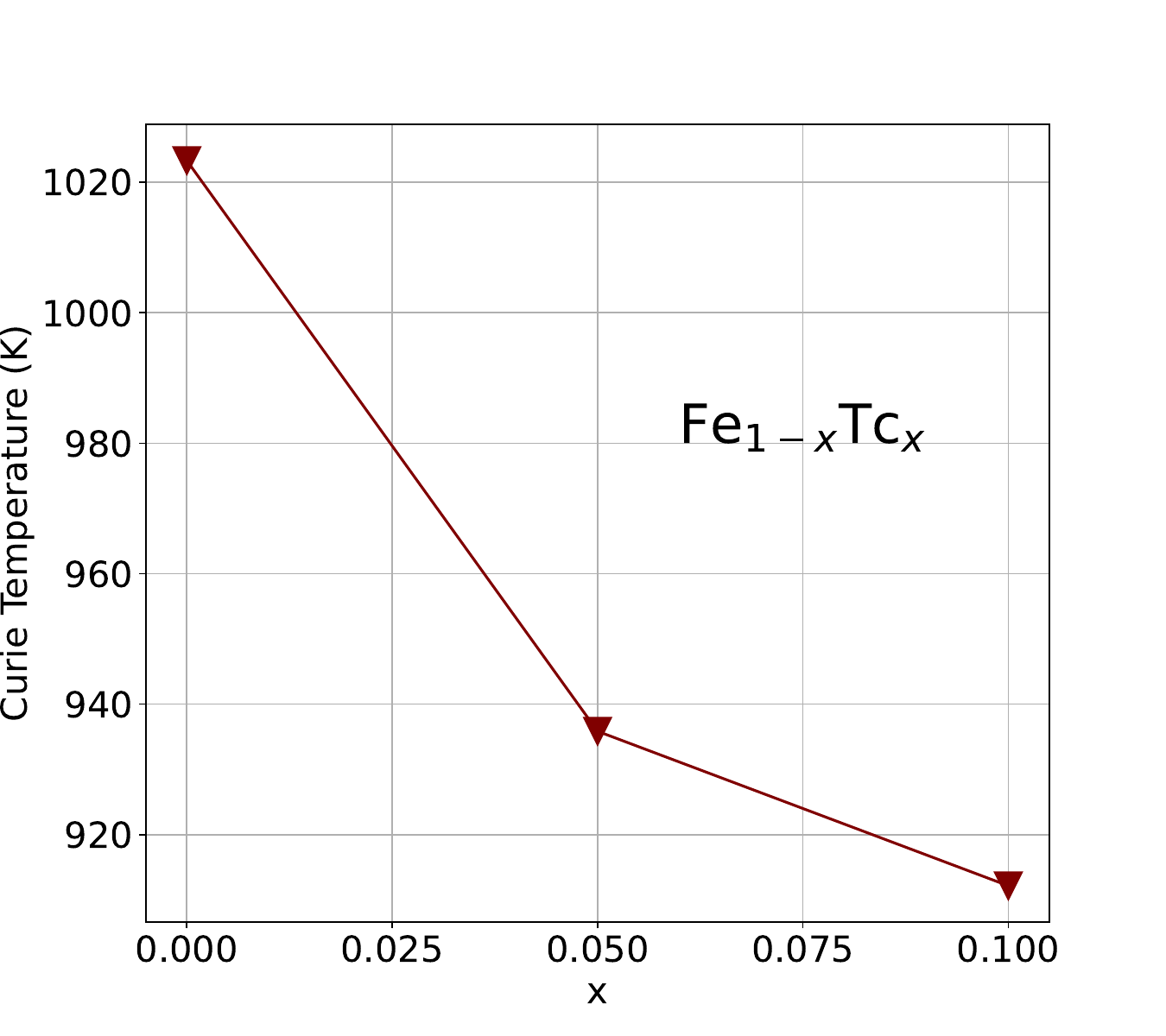}
    \caption{The predicted Curie temperatures of bcc Fe$_{1-x}$Tc$_x$ as a function of the fractional substitution of iron with technetium.}
    \label{fig:FeTc}
\end{figure}

As a final step, we use the model to predict the behavior of bcc Fe$_{1-x}$Tc$_x$ (iron-technetium). The experimental investigation of the magnetic phase diagram of this system remains unexplored due to the radioactive nature of technetium. This makes it a good candidate for a test that can be evaluated against future experiments. Figure \ref{fig:FeTc} shows the predicted $T_\text{C}$ for $x = 0, 0.05,$ and $0.10$. The technetium atoms have completely half-filled $4d$ orbitals and the magnetic moments after the ground-state search are approximately $0.5 \mu_B$. Hence, the technetium magnetic moments are left unconstrained in the DLM simulations. 


\section{Discussion \label{sec:discussion}}
 Our results show the model of Eq.~(\ref{eq:model_D_form}) to be generally capable of reproducing the qualitative effect on $T_\text{C}$ of altering the alloy composition. In all cases presented in Fig.~\ref{fig:all_alloys_composition}, the model accurately predicts the trends between composition and the Curie temperature, and in many cases, even a quantitative agreement is observed. 
 
 For the Fe$_{1-x}$V$_x$ system, the difference between the experimental and predicted values is at most $60$ K and at best around 20 K. The predicted $T_\text{C}$ do not quite follow the curve of the experimental $T_\text{C}$ which initially increases up to a max at around $x = 0.125$ and followed by a decrease in $T_\text{C}$. Instead, for the predicted $T_\text{C}$, a slight increase is estimated from pure iron to 5\% vanadium which flattens out, followed by yet another increase between $x =0.125$ to $x=0.25$. However, the change in $T_\text{C}$ for these alloys are minimal and our method correctly predicts this observation.
 
 In Fe$_{1-x}$Cr$_x$, the predicted $T_\text{C}$ are almost on top of the experimental values, with the exception of the highest  chromium content, $x = 0.375$, where there is a deviation of approximately 80 K. The curvature of the experimental $T_\text{C}$ is well estimated by our model. 
 
 For the Fe$_{1-x}$Co$_x$ system, our model shows especially good agreement in the cobalt-rich fcc region, where the deviation is at most 60 K. In the bcc region, the error stays below $80$ K.  Overall, the errors of the Fe-containing binary alloys are consistently within our previous estimate of the mean absolute error, 126 K, of the model \cite{brannvall_predicting_2024} and the trends with the alloy composition are well described. 

 As mentioned previously, the chromium and vanadium magnetic moments of Fe$_{1-x}$Cr$_x$ and Fe$_{1-x}$V$_x$ are in some cases of a magnitude greater than $0.75$ $\mu_B$ in the ground-state search. Hence, these moments should be constrained in the DLM run according to our general workflow. However, this was not done since this prevented the DFT calculations from achieving satisfactory electronic convergence in the self-consistent cycle. The energy difference between iterations remained in the order of 1 eV, well above the target threshold. This is attributed to the high level of itineracy of the chromium and vanadium moments in these binary alloys. The moments are very sensitive to the magnetic environment. Constraining the directions of all moments in a DLM state destabilizes these itinerant magnetic moments, making it challenging for the self-consistency cycle to converge. When running DLM calculations with unconstrained chromium and vanadium moments, their magnitudes decrease significantly, from $1.3$ $\mu_B$ to $0.2$ $\mu_B$ in the case of 12.5\% chromium. A similar shift is observed for the vanadium moments in all investigated compositions of Fe$_{1-x}$V$_x$. This can be compared with Fe$_{1-x}$Co$_x$ where the cobalt magnetic moments are borderline cases between local and itinerant, but are stabilized by the iron moments. 
 In this case, both the cobalt and iron moments are therefore constrained. 

 Our model successfully captures the concentration dependence of $T_\text{C}$ in Ni$_x$Cu$_{1-x}$MnSb, aligning with experimental observations. The error is generally within 150 K, with a maximum of approximately 170 K. This is comparable to the effects of 5\% intersite disorder in this half-Heusler alloy, as reported in Ref.~\cite{alling_effect_2009}, but not considered in the present work. The predicted $T_\text{C}$ values display a nearly linear decrease as copper content increases especially in the copper-rich region, with a slight underestimation compared to experimental data.
 For the Ni$_{x}$MnSb system, the predictions are very close to the experimental values, with the largest error being around 100 K at $x=1.75$. This is also in the composition range at which Ref.~\cite{webster_chemical_1984} observed increasing disorder form the C1$_b$ structure.
 
In Fig.~\ref{fig:all_alloys_composition}, a hint of an increase in $T_\text{C}$ as $x$ increases is seen for  the Ti$_{1-x}$Cr$_x$N system. The incline is comparable to the experimental incline, however, with a significant overestimation of the absolute value of $T_\text{C}$. The magnetic moments of the  chromium atoms did not align completely ferromagnetically in the ground-state search; the energies are, however, relatively similar to the ferromagnetic case. Consistently over the different $x$-values, the ground state resulted in a slightly lower energy compared to a perfect ferromagnetic configuration. The largest difference is for $x=0.5$, which is also the upper limit for ferromagnetism in this system \cite{aivazov_magnetic_1975,inumaru_ferromagnetic_2007,alling_theory_2010}. These differences do not change the predicted $T_\text{C}$ significantly. 

For Ti$_{1-x}$Cr$_x$N the experimental Curie temperatures are relatively low. One should note that the model of Eq.~(\ref{eq:model_D_form}) has a lower boundary of 124 K, since $\Delta E$ is never below zero. Approaching a Curie temperature near 124 K will therefore naturally lead to problems in our predictions. The experimental $T_\text{C}$ in Ti$_{1-x}$Cr$_x$N  is at most 140 K, which is very close to the lower limit of our model. This overestimation is therefore likely due to the inherent limitations of the model. A way of resolving this could be by introducing a correction term enabling differentiation between the low and high $T_\text{C}$ regions. The design of our model is, however, based on the essential physical assumptions described briefly in Sec. \ref{sec:rationale_model} and in more detail in Ref.~\cite{brannvall_predicting_2024}. Introducing a higher level of complexity into the model structure in an attempt to resolve this issue would potentially obscure the physics assumptions on which this model is based. 


 From Fig.~\ref{fig:CoAl}, it is clear that $T_\text{C}$ predictions of fcc Co$_{1-x}$Al$_x$ with our model are far from the experimental values. The steep decline with added aluminum content is not captured in the model until it is above 10\% aluminum. Although the qualitative effect of decreasing $T_\text{C}$ is reproduced, we attribute the lack of quantitative agreement to the cobalt magnetic moments being of a more itinerant type, which becomes especially apparent in the case of Co$_{1-x}$Al$_x$. From Fig.~\ref{magmom_Co}, it is seen that the magnetic moments of the cobalt atoms do not have significantly different magnitudes when comparing fcc Co$_{1-x}$Al$_x$ and fcc Co$_{1-x}$Fe$_x$. There are slightly more magnetic moments with magnitudes below 0.5 $\mu_B$ in the case of Co$_{1-x}$Al$_x$. The large iron moments, seen in Fig.~\ref{magmom_Fe-Al}, might instead be the reason for the fcc Co$_{1-x}$Fe$_x$ system being more stable and our model consequently giving better estimations in that case.

As the technetium content increases, $T_\text{C}$ decreases in bcc Fe$_{1-x}$Tc$_x$, as shown in Fig.~\ref{fig:FeTc}. This trend corresponds well to the behavior observed in experimentally studied systems such as Fe$_{1-x}$Ru$_x$ and Fe$_{1-x}$Mo$_x$, which also contain $4d$ elements that lower the Curie temperature \cite{pottker_mossbauer_2004, arajs_ferromagnetic_1965}. However, unlike technetium, these elements are not radioactive. Molybdenum, in particular, has a very similar electronic structure to technetium, with both elements having completely half-filled $4d$ orbitals, which suggests that technetium might contribute to the magnetic interactions in iron alloys in a way comparable to molybdenum. Similarly, Fe$_{1-x}$Re$_x$ exhibits a comparable electronic structure, but with $5d$ electrons. In this case, a reduction in the Curie temperature is also observed with increasing rhenium content \cite{cieslak_discovery_2014}.

\section{Conclusions}
In this work we have applied the model developed in our previous work Ref.~\cite{brannvall_predicting_2024} to the substitutional disordered alloys bcc and fcc Fe$_{1-x}$Co$_x$, bcc Fe$_{1-x}$Cr$_x$, bcc Fe$_{1-x}$V$_x$, Ni$_{x}$Cu$_{1-x}$MnSb, Ni$_x$MnSb, Ti$_{1-x}$Cr$_x$N, fcc Co$_{1-x}$Al$_x$ and the experimentally unexplored bcc Fe$_{1-x}$Tc$_x$. The model accurately predicts the trends of the variation of $T_\text{C}$ with composition and in some cases is in good absolute agreement with experimental values. Our model has a lower limit of 124 K which naturally becomes apparent for systems with low $T_\text{C}$ such as Ti$_{1-x}$Cr$_x$N. However, the model still predicts an accurate $\partial T_\text{C}/\partial x$ in this system. Systems with a more itinerant type of magnetism, e.g., Co$_{1-x}$Al$_x$, are not well described by this method, as it depends mainly on the rotational disorder of robust magnetic moments to describe the paramagnetic state. Overall, the approach provides a valuable and computationally efficient tool for predicting trends in $T_\text{C}$, offering insights that can guide experimental studies and alloy design. This approach not only increases computational efficiency compared to other theoretical methods but also allows for a broader applicability, without the need for extensive system-specific human intervention.

\section*{Acknowledgement}
The computations were enabled by resources provided by the National Academic Infrastructure for Supercomputing in Sweden (NAISS), partially funded by the Swedish Research Council through grant agreement no. 2022-06725. B.A. acknowledges financial support from the Swedish Research Council (VR) through Grant No. 2019-05403, and 2023-05194 and from the Swedish Government Strategic Research Area in Materials Science on Functional Materials at Linköping University (Faculty Grant SFOMatLiU No. 2009-00971). R.A. acknowledges financial
support from the Swedish Research Council (VR)
Grant No. 2020-05402 and the Swedish e-Science Research
Centre (SeRC).

\appendix
\section{Numerical values of predicted and \\experimental $T_\text{C}$\label{appendix_A}}
\begin{table}[H]
\caption{Table with predicted and experimental $T_\text{C}$ for the systems of Fig. \ref{fig:all_alloys_composition}.}
\centering
\begin{tabular}{|c|c|c|c|}
\hline
\textbf{System} & \textbf{x-value} & \textbf{$T_\text{C}^\text{Model}$ (K)} &  \textbf{ $T_\text{C}^\text{Exp.}$ (K)}\\
\hline
\multirow{3}{*}{$\mathrm{Fe_{1-x}Co_x}$} 
    & 0.0 & 1023 & 1043\\
    & 0.125 & 1080 & 1064\\
    & 0.25 & 1151 & 1230\\
    & 0.375 & 1293 & - \\
    & 0.75 & 1143 & 1178 \\
    & 0.9 & 1365 & 1308 \\
    & 1.0 & 1418 &  1388\\
\hline
\multirow{3}{*}{$\mathrm{Fe_{1-x}V_x}$} 
    & 0.05 & 1062 & 1083\\
    & 0.125 & 1053 & 1111\\
    & 0.25 & 1091 & 1053\\
\hline
\multirow{3}{*}{$\mathrm{Fe_{1-x}Cr_x}$} 
    & 0.1 & 1049 & 1023\\
    & 0.125 & 1032 & 1010\\
    & 0.25 & 919 & 906\\
    & 0.375 & 847 & 767 \\
\hline
\multirow{3}{*}{$\mathrm{Ni_xCu_{1-x}MnSb}$} 
    & 0.25 & 145 & 300 \\
    & 0.375 & 246 &  400\\
    & 0.5 & 370 &  541\\
    & 0.675 & 497 &  640\\
    & 0.75 & 518 & 670 \\
    & 0.875 & 634 & 720 \\
    & 1.0 & 663 & 755 \\
\hline
\multirow{3}{*}{$\mathrm{Ni_xMnSb}$} 
    & 1.25 & 488 & 584 \\
    & 1.5 & 408 &  470\\
    & 1.75 & 332 & 438 \\

\hline
\multirow{3}{*}{$\mathrm{Ti_{1-x}Cr_x N}$} 
    & 0.125 & 189 & - \\
    & 0.25 & 224 &  50\\
    & 0.375 & 247 & 100 \\
    & 0.5 & 207 & 140 \\

\hline
\end{tabular}
\label{tab:alloys_numerical_values}
\end{table}
\section*{Data availability}
Python scripts used to extract the data from density functional theory calculations are openly available in the DFT\_based\_CurieTemp\_predictions repository at \href{https://github.com/marbr639/DFT_based_CurieTemp_predictions}{https://github.com/marbr639/DFT\_based\_CurieTemp\\\_predictions}. Other data that support the findings of this study are available upon request.







\end{document}